\documentclass[10pt]{iopart}
\usepackage{graphicx}
\usepackage{hyperref}

\begin{document}
	
	\title[Direct implanted Back-gates developed for GaAs heterostructures]{Donor implanted Back-gates in GaAs for MBE-grown highest mobility two-dimensional electron systems}
	
	\author{J. Scharnetzky$^1$, P. Baumann$^1$, C. Reichl$^1$, H. Karl$^2$, W. Dietsche$^1$, and W. Wegscheider$^1$}
	
	\address{$^1$ Solid State Physics Laboratory, ETH Z\"urich, 8093 Z\"urich, Switzerland}
	\address{$^2$ Lehrstuhl für Experimentalphysik IV, Universität Augsburg, 86159 Augsburg, Germany}
	
	\ead{janscha@phys.ethz.ch}
	\vspace{10pt}
	\begin{indented}
		\item[]March 2021
	\end{indented}
	
	\begin{abstract}
		Three different elements, Silicon, Selenium, and Tellurium, are ion-implanted in Gallium arsenide to form a conducting layer that serves as a back-gate to a molecular beam epitaxy (MBE) overgrown two-dimensional electron gas (2DEG). While the heavy ion Tellurium creates too many defects in the gallium arsenide to form a conducting layer, both Silicon and Selenium show promising results combined with MBE-grown high-quality 2DEGs. Similar 2DEG mobilities compared to non-implanted reference samples are achieved for both Silicon and Selenium implanted structures. Individual contacts to the back-gate are challenging. However,  Silicon implanted structures, annealed before the MBE growth, result in a functional back-gate, and the electron density of the 2DEG can be tuned via the back-gate.
	\end{abstract}
	
	%
	%
	%
	%
	%
	
	\section{Introduction}
	Various experiments demonstrate the requirement for electric top- and back-gates combined with highest mobility two-dimensional electron gases \cite{Roeoesli2020, Yoon2010, Scharnetzky2020}. Metallic top-gates can be prepared easily after the MBE growth using standard photolithography and metal deposition. The fabrication of non-global back-gates is more complex as they need to be processed before the MBE growth. However, any modification of the surface typically has a deteriorating effect on the quality of the overgrown 2DEG. Therefore, the challenge is to develop a residue-free process, which has a minimal impact on the surface quality to allow the growth of crystalline, defect-free epitaxial layers.
	
	Previous attempts creating patterned back-gates used a metal-organic vapor deposition grown (MOCVD) highly n-doped GaAs layer, which is then chemically wet etched to create shallow edges of the conducting back-gates \cite{Rubel1998}. Overgrowing these shallow etched samples with a heterostructure resulted in reliability problems with strongly varying 2DEG qualities across the shallow edges.  Also, during operation in high magnetic fields,  the contacts to the 2DEG often become insulating. 
	
	Ion implantation fulfills all the requirements to fabricate back-gates for high mobility applications. The technique is planar, and for sufficient implantation energies, it does not significantly modify the gallium arsenide surface as most dopants and their energy are distributed well underneath the surface. To this end, our group recently introduced this technique to fabricate electric back-gates \cite{Berl2016}. In this procedure, the patterned back-gates are fabricated by oxygen ion implantation into a highly Silicon doped gallium arsenide layer. This layer, grown by MOCVD on a semi-insulating GaAs substrate, has been patterned with photoresist before the implantation. In regions not covered by photoresist, oxygen ions passivate the n-doped layer. Covered areas remain conductive as the implanted ions stop within the resist.
	
\begin{figure}[h]
		\centering
		\includegraphics[width=0.8\textwidth]{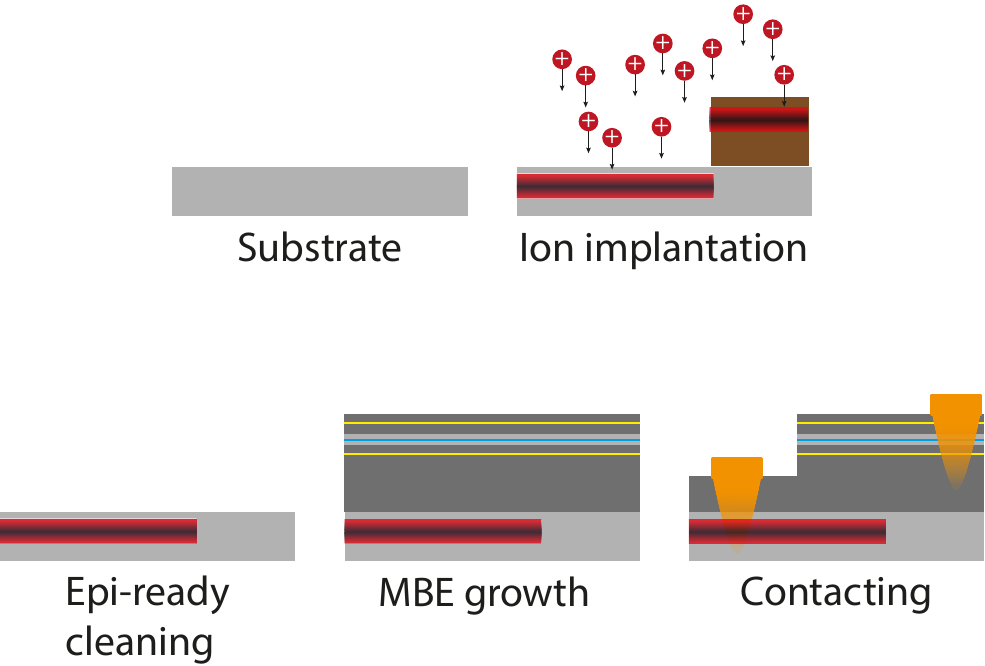}
		\caption{Process to fabricate structured ion-implanted back-gates: A photo resist-patterned gallium arsenide substrate is ion-implanted using Selenium, Tellurium, or Silicon. Before loading the sample into the MBE chamber, the surface is cleaned using the epi-ready process to remove all residues. The ion-implanted substrate is overgrown with a heterostructure. The blue line marks the 2DEG, and the two yellow lines the Silicon dopings. In a final step, the 2DEG and back-gate are contacted separately using photolithography and chemical wet-etching.}
		\label{fig:schema}
	\end{figure}

	A significant simplification is the direct implantation of dopants to create conducting back-gates. Figure \ref{fig:schema} indicates the process, where the implanted ions produce the conductive region. The major challenge is the dopant activation without modifying the substrate's surface. To obtain a high yield in electrical activation of the implanted donors, rapid thermal annealing (RTA) using temperatures exceeding 900$^{\circ}$C is required \cite{Jacobson1985}. To prevent outgassing and thus modification of the crystalline surface, encapsulation, typically using silicon nitride is required \cite{Davies1985}. However, to minimize the MBE chamber's contamination, the dopants' annealing and activation step preferably occurs in-situ during MBE growth. Typical MBE growth temperatures are around 650$^{\circ}$C, and the chamber's arsenic pressure protects the GaAs wafer's surface. However, the reduced temperatures limit the dopant's activation.

	This publication focuses on three ion species as potential donors in GaAs: Selenium, Tellurium, and Silicon. The implantation energies are chosen in such a way that the peak ion concentration is located well below the surface of the GaAs substrate. Silicon, a typical donor dopant in the gallium arsenide technology,  has the drawback that it may be incorporated in an amphoteric way depending on the annealing temperature \cite{PEARTON1993}. Additionally,  Selenium and Tellurium, which are n-dopants in GaAs independently of their GaAs lattice sites, are tested.

	\section{Experimental details}
	Three different experiments are performed to determine the ideal implantation parameters for patterned back-gates:
	Firstly solely the activation of the implanted ions is characterized without MBE growth. Secondly, implantation parameters that result in sufficiently conducting layers are combined with an MBE overgrowth, without patterned photoresist, to test the implantation's impact on the 2DEG quality. The whole process to fabricate patterned back-gates is performed in a third step as described in figure \ref{fig:schema}.
	
	For the characterization test, semi-insulating GaAs wafer pieces without prior photolithography are ion-implanted using an NV 3206 System from Axcelis Technology at University of Augsburg. Table \ref{tab:imppara} lists the implantation parameters for Silicon, Tellurium, and Selenium. Silicon and Selenium are single ionized, while tellurium is double ionized for implantation. The projected range has been calculated using SRIM simulations \cite{Ziegler2010}. All implantation energies are selected to bury the peak concentration of dopants well in the substrate to preserve the surface, which is paramount for high-quality MBE growth. For Selenium and Tellurium four and for Silicon three, different doses are implanted in the range indicated in table \ref{tab:imppara}.
	
	\begin{table}[h]
		\caption{\label{tab:imppara}The different implantation parameters for the three types of ions.  The depth is calculated using TRIM simulations, see text for details.}
		\begin{indented}
			\item[]\begin{tabular}{@{}llll}
				\br
				Implant type &Energy &Dose& Projected range  \\
				&(keV) &(cm$^{-2}$)&(nm)\\
				\mr
				Silicon   &90 &1$\times$ 10$^{12}$ to 1  $\times$ 10$^{14}$&86\\
				Selenium  &200&1$\times$ 10$^{13}$ to 5  $\times$ 10$^{14}$&81\\
				Tellurium &300&1$\times$ 10$^{13}$ to 5  $\times$ 10$^{14}$&86\\
				
				\br
			\end{tabular}
		\end{indented}
	\end{table}
	
	For characterization of the dopant's activation, the samples are annealed in a rapid thermal annealer (Rapid Thermal Processor AS-One 100) at 700 $^{\circ}$C between 30 and 120 seconds in an H2/N2 ambient. To protect the samples' surface during annealing, they are covered with another GaAs wafer piece face-to-face \cite{Pearson1993}. The samples are then contacted by soldering indium contacts and subsequently furnace annealed at 450$^{\circ}$C for 10 minutes. Afterwards, using the van-der-Pauw method, square resistance, density, and mobility of the samples are measured at room temperature and 4 Kelvin in a liquid helium dipstick system. 
	
	Implantation parameters that resulted in sufficient conductivity of the GaAs are then tested in the MBE system. Activation of the implanted ions occurs during the growth of a 2DEG heterostructure. Simultaneously, a reference sample in this case on a non-implanted GaAs piece is grown. The samples are contacted again by soldering Indium with subsequent annealing. The back-gate and the 2DEG are contacted in parallel, which allows determining the 2DEG quality on top of the implanted region and the presence of the conducting back-gate. During magnetotransport measurements, the back-gate manifests itself as a parallel channel that is not present in the simultaneously grown reference sample. The longitudinal resistance will not reach zero Ohm at integer quantum hall states but will remain larger due to the back-gate's finite resistance.
	Additionally, diffusion of the implanted dopants during the long-time growth process in the MBE system (up to 5 hours) is tested using SIMS measurements of the overgrown samples.
	
	In the final experiment, patterned back-gates defined by photoresist and ion implantation are fabricated with the selected implantation parameters. The patterned back-gates allow contacting the back-gate and the 2DEG separately. After the epi-ready cleaning process to remove all residues from the processing \cite{Berl2016}, the samples are overgrown with a high mobility 2DEG heterostructure. The buffer structure, the region between the back-gate and the 2DEG, consists of an an AlGaAs alloy with high Aluminum content to prevent electrical shorts between the back-gate and the 2DEG \cite{Scharnetzky2020}. A Hall-bar of dimensions 650 $\mu$m $\times$ 200 $\mu$m is prepared using standard photolithography and wet etching. An Au/Ge/Ni alloy is deposited and annealed to contact both back-gate and 2DEG before the samples are measured in a helium dipstick system at 4 and 1 K. 
	
	\section{Results}
	The results from the three different characterization measurements explained in the previous section are presented. 
	\subsection{Characterization of the implanted and RTA-annealed samples}
	In table \ref{tab:char} the electrical characterization at 4 K after RTA annealing at 700 $^{\circ}$C between 30 and 120 seconds is shown. None of the tellurium implanted samples could be activated. Tellurium, the heaviest tested element, requires higher energy during the implantation to reach the desired depth profile. It is likely that the resulting high number of defects created during its implantation and the relatively low annealing temperature is not sufficient to create a conductive layer. Therefore, tellurium is discarded as implantation material.  
	
	For selenium, doses between 5 $\times$10$^{13}$ cm$^{-2}$ and 5 $\times$10$^{14}$ cm$^{-2}$ show sufficient and reliable conductance for the tested samples. The lowest dose, 1 $\times$10$^{13}$cm$^{-2}$, shows large variations in two-point resistances, and therefore it has been discarded for further tests. 
	
	Similar to Selenium, the lowest Silicon dose does not result in a conducting layer after thermal annealing. The doses 1 $\times$10$^{13}$cm$^{-2}$ and 1 $\times$10$^{14}$cm$^{-2}$ result in conducting layers after thermal annealing. The low number of defects created during the implantation results in a higher electron mobility than for the Selenium implantation.

	\begin{table}[h]
		\centering 
		\caption{\label{tab:char}The electrical characterization of Silicon (Si), Selenium (Se) and Tellurium (Te) implanted and annealed samples are shown, measured at 4 K.}
		\begin{indented}
			\item[]\begin{tabular}{@{}llll}
				\br
				Implant type and dose&Two-Point Resistances&Density &Mobility \\
				(cm$^{-2}$) &(k$\Omega$) &(cm$^{-2}$)&(cm$^2$V$^{-1}$s$^{-1}$)\\
				\mr
				Si 1 $\times$ 10$^{12}$&not measurable&&\\
				Si 1 $\times$ 10$^{13}$&1.7 - 2.1&5.4 $\times$ 10$^{11}$&1990\\
				Si 1 $\times$ 10$^{14}$&1.2 - 1.3&1.0 $\times$ 10$^{13}$&1480\\
				Se 1 $\times$ 10$^{13}$&29 - 89&6.1 $\times$ 10$^{11}$&530\\
				Se 5 $\times$ 10$^{13}$&52 - 61&6.3 $\times$ 10$^{11}$&250\\
				Se 1 $\times$ 10$^{14}$&15 - 16&9.0 $\times$ 10$^{11}$&540\\
				Se 5 $\times$ 10$^{14}$&4.0 - 4.9&2.4 $\times$ 10$^{12}$&760\\
				Te all doses		   &not measurable&&\\
				\br
			\end{tabular}
		\end{indented}
		
	\end{table}

	\subsection{MBE compatibility of implanted back-gates}
	Samples implanted with the three selected selenium doses, between 5$\times 10^{13}$ to 5$\times 10^{14}$cm$^{-2}$, are cleaned using the epi-ready cleaning process prior to the MBE loading. Three samples are mounted next to each other on a supporting wafer, which serves as a reference during the growth. 
	
	As described before we use a high aluminum concentration in the buffer to suppress diffusion and create electric isolation between back-gate and 2DEG. Towards the 2DEG, the aluminum concentration is reduced to a typical value of 30$\%$ for 2DEG growth.

\begin{figure}[h]
		\centering
		\includegraphics[width=0.8\textwidth]{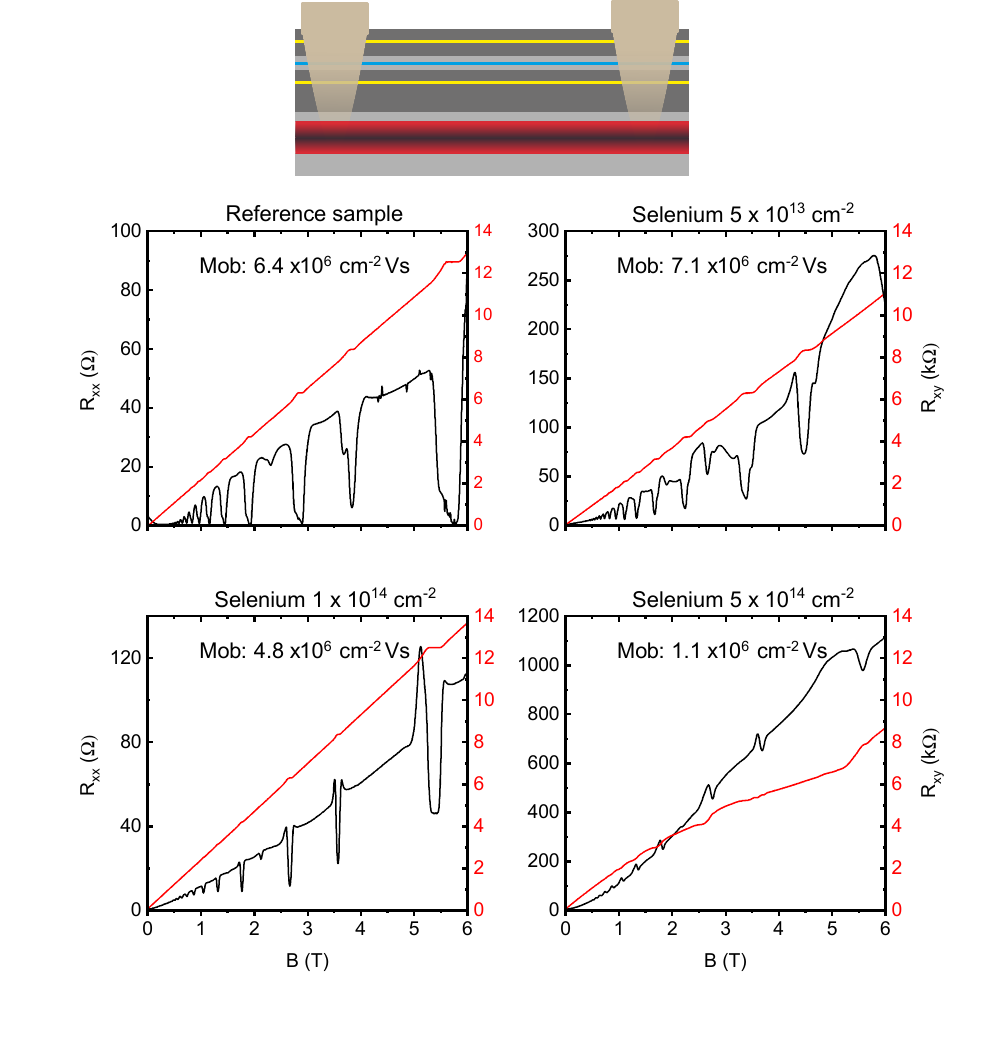}
		\caption{Top: A sketch of the parallel contacted back-gate (in red) and 2DEG (in blue) by soldering Indium. Bottom: The Shubnikov-de-Haas (black) and Hall traces (grey) are shown for samples with three different implantation doses, as well as for a simultaneously grown reference sample, all measured at 1.3 K.}
		\label{fig:magtrans}
	\end{figure}

	In table \ref{tab:charReg1K}, the characterization results are shown at 1.3 K.  For the 5 $\times$ 10$^{13}$ and the 1 $\times$ 10$^{14}$cm$^{-2}$ doses, the quality of the 2DEG of the implanted substrates is similar to the reference sample's quality. The 5 $\times$ 10$^{14}$cm$^{-2}$ dose shows a reduced mobility. Overall, the samples' 2DEGs are of high quality with a mobility above 1 $\times 10^{6}$cm$^2$ V$^{-1}$ s$^{-1}$. Magnetotransport measurements shown in figure \ref{fig:magtrans}, taken at 1 K, verify the high 2DEG quality. The Hall resistance trace shows quantized plateaus for all doses but the 5 $\times$ 10$^{14}$cm$^{-2}$ dose. The strong parallel channel conceals the quantized plateaus in the Hall trace. 
	
	Regarding the longitudinal resistance trace of the reference sample, the minima approach 0 $\Omega$. Observing the similar Shubnikov-de-Haas traces for the implanted samples, non-zero minima for quantized Hall plateaus can be observed, indicating a parallel channel. For increasing implantation doses, the parallel channel is more pronounced. The origin is the parallel contacted back-gate, as otherwise, the samples are identical to the reference structure.

	\begin{table}[h]
		\caption{\label{tab:charReg1K}The Selenium implanted structures overgrown with a high mobility heterostructure, are characterized without illumination  at 1.3 K. The density is derived from Hall measurements.}
		\begin{indented}
			\item[]\begin{tabular}{@{}llll}
				\br
				&sheet resistance  & density & mobility \\
				&($\Omega$)& (10$^{11}$cm$^{-2}$)&(10$^6$ cm$^2$ V$^{-1}$ s$^{-1}$)\\
				\mr
				Ref. sample &3.5  & 2.8&6.4  \\
				Se 5$\times$10$^{13}$&  3.1&3.4&7.1\\
				Se 1$\times$10$^{14}$& 	4.9&2.7&4.8\\
				Se 5$\times$10$^{14}$& 21&2.9&1.1 \\
				
				\br
			\end{tabular}
		\end{indented}
	\end{table}
	
\begin{figure}[h]
		\centering
		\includegraphics[width=0.8\textwidth]{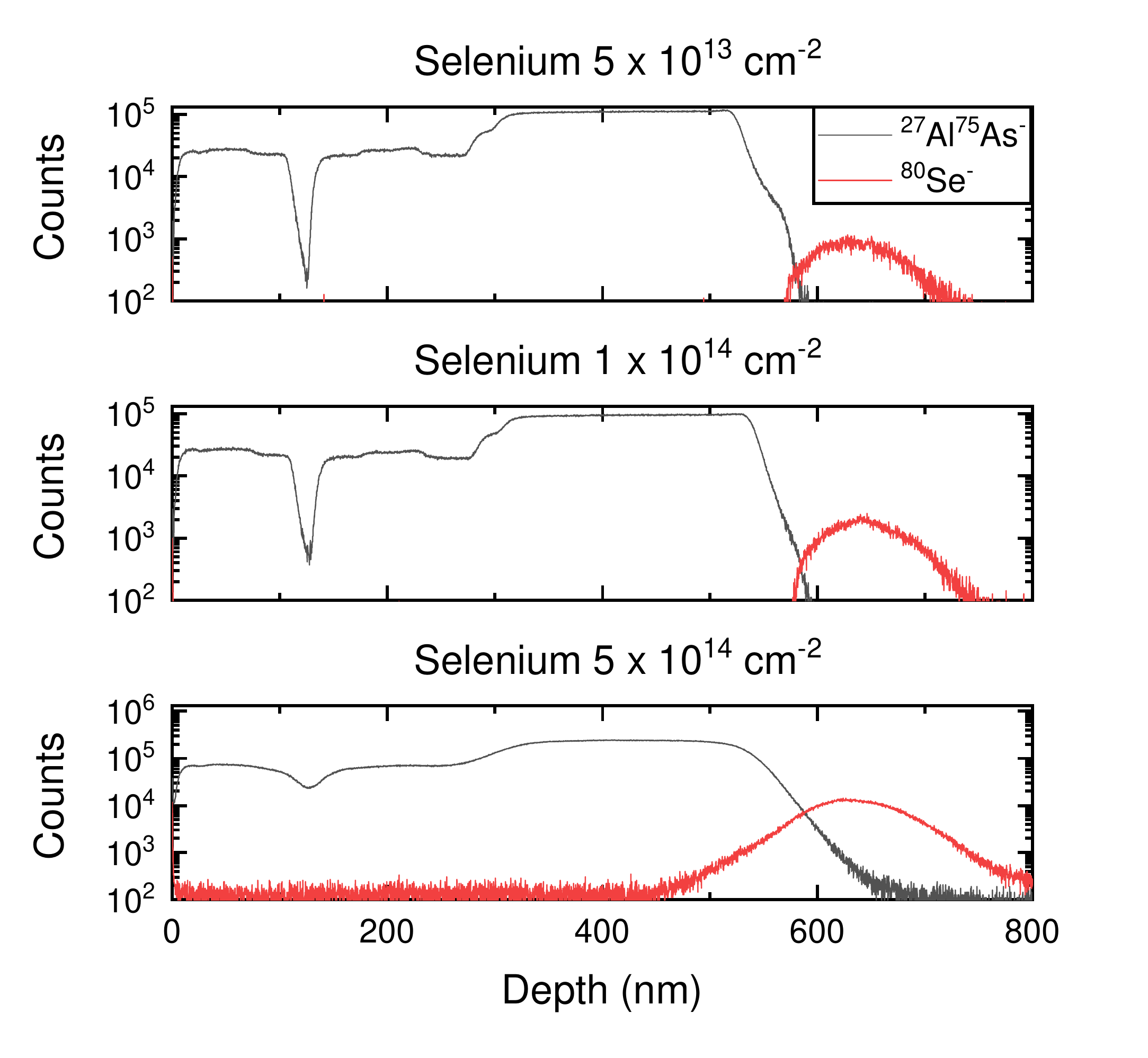}
		\caption{SIMS measurements for the three tested implantation doses are shown, the substrate starts around 600 nm below the surface, it can be well recognized by the strong decrease in the Aluminum signal. The buffer structures clearly show the two different aluminum concentrations used, the high ratio close to the substrate and the reduced content close to the quantum well (QW). The minima in Aluminum concentration identifies the QW location.}
		\label{fig:SIMS}
	\end{figure}

	To analyze the effect of the long-term in-situ MBE annealing on the implantation, SIMS measurements have been performed on the implanted and overgrown samples, shown in figure \ref{fig:SIMS}. For all the implantation doses but the 5 $\times$ 10$^{14}$cm$^{-2}$ dose, the Selenium implanted ions do not diffuse into the MBE-grown structure. The peak value is at the expected position. On the other hand, the 5 $\times$ 10$^{14}$cm$^{-2}$ dose sample shows strong diffusion, which explains the reduced 2DEG quality observed in the characterization measurements. Similar enhanced diffusion for high implantation doses has been reported previously \cite{Krishnamoorthy1998}. 
	
	The compatibility with high-quality MBE growth is also proven for the Silicon implantation. The 2DEG mobility of the implanted samples is comparable with the reference structures, as shown in table \ref{tab:charSi1K}. Also, no diffusion of the implanted ions is observed during SIMS characterizations for the selected implantation doses. However, no parallel conducting channel is detectable during magnetotransport measurements of a parallel contacted back-gate and 2DEG. A possible reason is that the Indium contacts do not sufficiently contact the back-gate. Also, a lack of or an amphoteric activation of the silicon ions during the MBE growth may be the reason. 
	
	\begin{table}[h]
		\caption{\label{tab:charSi1K}The Silicon implanted structures overgrown with a high mobility heterostructure, are characterized without illumination  at 1.3 K. The density is derived from Hall measurements.}
		\begin{indented}
			\item[]\begin{tabular}{@{}llll}
				\br
				&sheet resistance  & density & mobility \\
				&($\Omega$)& (10$^{11}$cm$^{-2}$)&(10$^6$ cm$^2$ V$^{-1}$ s$^{-1}$)\\
				\mr
				Ref. sample &1.3  & 2.7&18.5  \\
				Si 1$\times$10$^{13}$&  1.4&2.8&15.7\\\mr
				Ref. sample & 1.3 & 3.6 & 13.4\\
				Si 5$\times$10$^{13}$& 	1.1&3.2&17.3\\

				\br
			\end{tabular}
		\end{indented}
	\end{table}
	
	Nevertheless, for both Selenium 5$\times$10$^{13}$ cm$^{-2}$, 1$\times$10$^{14}$ cm$^{-2}$ and both Silicon 1$\times$10$^{13}$ cm$^{-2}$, 5$\times$10$^{13}$ cm$^{-2}$ implantation doses, the process to fabricate structured back-gates with individual contacts to back-gate and 2DEG is tested.
	
	\subsection{Patterned back-gates to separate the contacts to back-gate and 2DEG}
	The full fabrication process described in figure \ref{fig:schema} is executed in a final step. However, for both the Selenium and Silicon implanted back-gates, it turned out to be impossible to reliably contact the back-gate using a Gold-Germanium-Nickel alloy.  
	
\begin{figure}[h]
		\centering
		\includegraphics[width=0.8\textwidth]{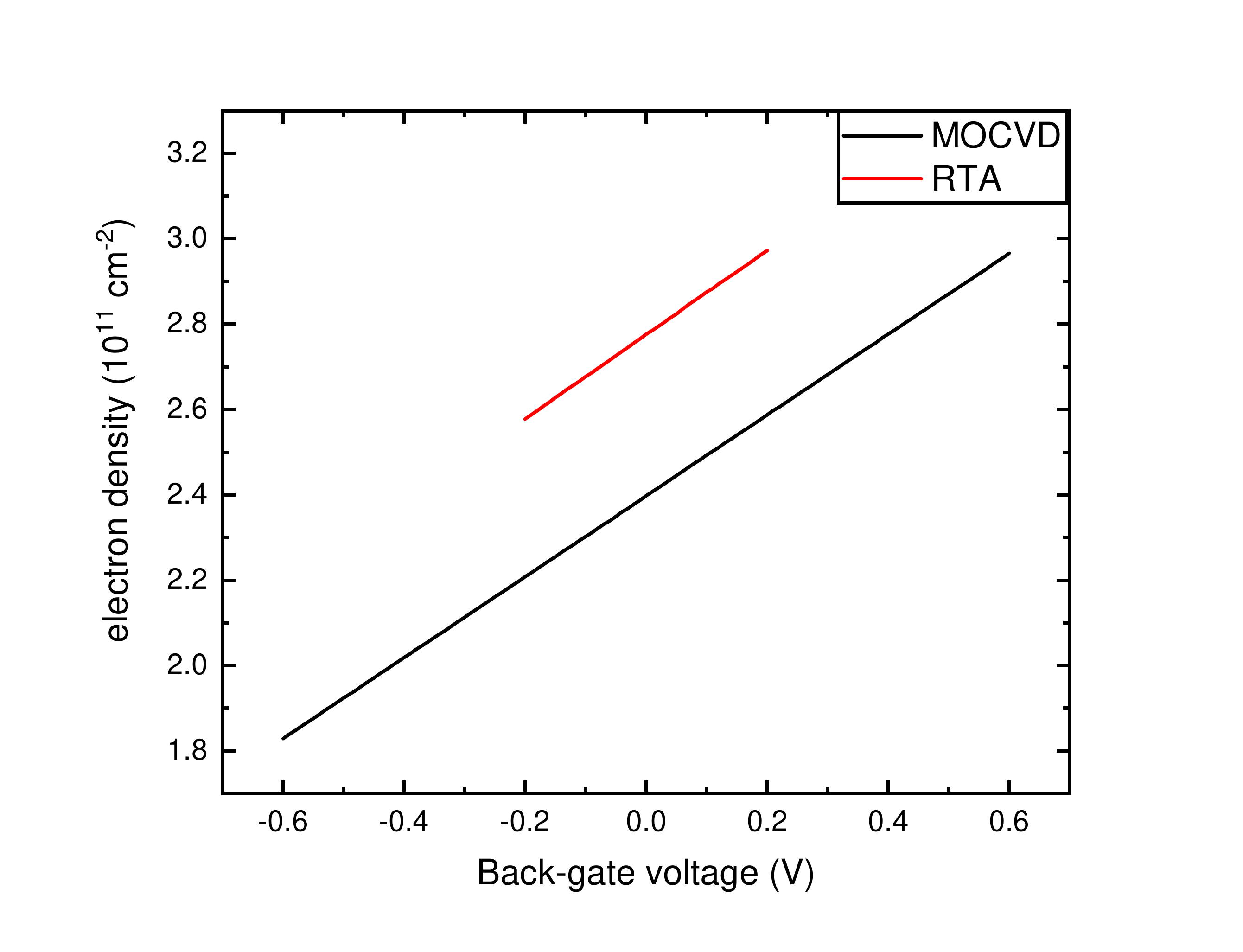}
		\caption{Density change of 2DEG using Si back-gates, preannealed using RTA/MOVCD systems}
		\label{fig:DensityRange}
	\end{figure}

	Regarding the Silicon implantation, already the  structures with parallel contacting of back-gate and 2DEG indicated a lack or amphoteric activation of implanted ions during the MBE growth. Therefore, an additional activation step has been introduced. The annealing step is tested using an RTA system at 600 $^\circ$C for 30 seconds and in an MOCVD system at 630 $^\circ$C for 10 minutes. During the annealing, the temperatures are intentionally set lower than at the pretests to avoid outgassing of the surface, especially for the RTA process. The epi-ready process is performed before the annealing to remove all the residues and prevent incorporation into the GaAs substrate during annealing. Annealing in the MOCVD system offers an intrinsically large Arsenic counter-pressure and no “face-to-face” wafer protection is needed. 
	In figure \ref{fig:DensityRange}, the 2DEG density is plotted as a function of back-gate voltage for the pre-annealed samples. The density to gate voltage characteristics is linear. The gate-range is reasonable large for the MOCVD-annealed sample but limited if annealing takes place in the RTA. Leakage between the back-gate and the 2DEG occurs beyond the plotted range. Additionally, the pre-annealing step drastically reduced the 2DEG quality, especially for the RTA annealed structure. Further optimization of the MOCVD-annealed structures should lead to a high-quality 2DEG with a functional Silicon implanted back-gate.

	\section{Conclusion}
	To summarize, ion implantation is compatible with the MBE-growth of high-quality heterostructures. The excellent quality of the 2DEG is demonstrated for Selenium and Silicon implanted structures and it decreases only slightly with increasing implantation doses. The fabrication of functional back-gates using the implantation of dopants remains challenging for the following reasons: 
	
	Silicon implants require an additional annealing step before the MBE-growth to form functional back-gates. The annealing process, either performed in an RTA or MOCVD system, reduced both the mobility of the 2DEG and the tuning range of the gate. 
	
	Tellurium implants require a higher temperature than acceptable within the MBE-system and are thus discarded as an implant material.
	
	Selenium implants show promising results during testing, such as a parallel conducting channel when contacting the implanted region in parallel with the 2DEG. However, so far, it has not been possible to fabricate functional back-gates based on selenium implantation. Reasons are likely, either an insufficient activation of dopants during the heat treatment within the MBE system or problems to contact the implanted layer using Au/Ge/Ni contacts. Further studies would be required to analyze this in detail.

	\section{Acknowledgment}
	We acknowledge the support through the Swiss National Foundation (SNF) and the NCCR QSIT (National Center of Competence in Research - Quantum Science and Technology). We are thankful to E. Gini for MOCVD annealing treatment performed at FIRST - Center for Micro- and Nanoscience, ETH Zurich. 
	\newpage
	\bibliography{Ionimplantation}
	\bibliographystyle{unsrt}
\end{document}